\documentclass{emulateapj}
\usepackage{times, graphicx}

\slugcomment{}

\shorttitle{Running Penumbral Waves in the Solar Chromosphere}
\shortauthors{D.B. Jess et al.}

\begin{document}

\title{The Influence of the Magnetic Field on Running Penumbral 
Waves in the Solar Chromosphere}

\author{D. B. Jess$^{1,2}$, V. E. Reznikova$^{1}$, T. Van Doorsselaere$^{1}$, 
P. H. Keys$^{2,3}$, D. H. Mackay$^{4}$}
\affil{$^{1}$Center for Mathematical Plasma Astrophysics, 
Department of Mathematics, KU Leuven,
Celestijnenlaan 200B bus 2400, B-3001 Heverlee, Belgium}
\affil{$^{2}$Astrophysics Research Centre, School of Mathematics and Physics, 
Queen's University Belfast, Belfast BT7 1NN, UK}
\affil{$^{3}$Solar Physics and Space Plasma Research Centre (SP$^{2}$RC), 
University of Sheffield, Hicks Building, Hounsfield Road, Sheffield S3 7RH, UK}
\affil{$^{4}$School of Mathematics and Statistics, University of St Andrews, St Andrews, 
Scotland, KY16 9SS, UK}
\email{d.jess@qub.ac.uk}

\begin{abstract}
We use images of high spatial and temporal resolution, obtained 
using both ground- and space-based instrumentation, to 
investigate the role magnetic field inclination angles play in the propagation 
characteristics of 
running penumbral waves in the solar chromosphere. Analysis of 
a near-circular sunspot, close to the center of the solar disk, reveals a smooth 
rise in oscillatory period as a function of distance from the umbral 
barycenter. However, in one directional quadrant, corresponding to 
the north direction, a pronounced 
kink in the period--distance diagram is found. 
Utilizing a combination of the inversion of magnetic Stokes vectors and 
force-free field extrapolations, we attribute this behaviour 
to the cut-off frequency imposed by the magnetic field geometry in this location. 
A rapid, localised inclination of the magnetic field lines in the north 
direction results in a faster increase in the dominant periodicity due to 
an accelerated reduction in the cut-off frequency.  
For the first time we reveal how the 
spatial distribution of dominant wave periods, obtained with one of the 
highest resolution solar instruments currently available, directly 
reflects the magnetic geometry 
of the underlying sunspot, thus opening up a wealth of 
possibilities in future magneto-hydrodynamic seismology 
studies. In addition, the intrinsic relationships 
we find between the underlying magnetic field geometries connecting the photosphere 
to the chromosphere, and the characteristics of running penumbral waves 
observed in the upper chromosphere, directly supports 
the interpretation that running penumbral wave phenomena 
are the chromospheric signature of 
upwardly-propagating magneto-acoustic waves generated in the photosphere. 
\end{abstract}


\section{Introduction}
\label{intro}

Waves and oscillations manifesting in the immediate vicinity of sunspots have 
been known for over $40$~years \citep{Bec69}. Early work on 
oscillatory phenomena in sunspot structures helped validate the detection 
of long-period oscillations, which are generated by the response of the 
umbral photosphere to the 5-minute $p$-mode global oscillations \citep{Tho82, Lit92}. 
While oscillations in solar active regions are dominated by periodicities 
intrinsically linked to the global $p$-mode spectrum, a wealth of alternative 
wave periods can also be identified in the sunspot locality, spanning three 
orders-of-magnitude from several seconds \citep{Jes07}, through to 
in-excess of one hour \citep{Dem85}.

The first observational evidence of running penumbral waves (RPWs) came from 
\citet{Giov72} and \citet{Zir72}, who detected concentric intensity waves propagating 
outwards through the penumbra of a sunspot. These waves, deemed as 
acoustic modes, were observed to propagate with a phase velocity of 
$10 - 20$~km{\,}s$^{-1}$ and intensity fluctuations in the range 
$10 - 20$\%. \citet{Bris97} and \citet{Kob04} have since revealed how 
the frequencies and phase speeds of RPWs are largest 
(3~mHz, 40~km{\,}s$^{-1}$) at the inner penumbral boundary, 
decreasing to their lowest values (1~mHz, 10~km{\,}s$^{-1}$) 
at the outer penumbral edge. Additionally, \citet{Kob00} has shown 
that the propagation of RPWs can be observed 
in the chromosphere 
up to 
$\sim$$15${\arcsec} ($\sim$$10{\,}000$~km) from the outer edge of the 
penumbral boundary, suggesting the quiet-Sun $p$-mode oscillations dominate at 
greater distances, hence over-powering the signatures of any remaining RPWs. 

The origin of RPWs has been under intense debate ever since their 
discovery, with current research attempting to address whether they are 
trans-sunspot waves of purely chromospheric origin 
\citep[e.g.,][and the references therein]{Tzio06, Tzio07}, or the 
chromospheric signature of upwardly-propagating $p$-mode 
waves \citep{Chri00, Chri01, Geor00, Cen06}. 
The sequence of studies by \citet{Chri00, Chri01} and \citet{Geor00} 
employed multi-height imaging and Doppler 
velocity measurements to examine the phase lag between 
sunspot oscillations at different layers of the lower solar 
atmosphere. 
Through examination of two-dimensional power maps and their 
resulting spatial coherence, 
the authors interpreted the correlation between 
umbral oscillations at various atmospheric heights and simultaneous 
RPWs as a strong indication that RPWs are excited by upwardly-propagating 
photospheric umbral oscillations. 
More recently, \citet{Tzio06, Tzio07} 
employed similar multi-height imaging and spectroscopy to 
investigate the coupling between umbral and penumbral oscillations 
using spectral, phase-difference, and coherence analysis techniques. 
However, the authors detected 
large jumps in the oscillation period and the intensity-velocity 
phase difference at the umbra-penumbra boundary, and as a result, 
were unable to convincingly support the upwardly-propagating 
$p$-mode wave scenario. 
Conversely, \citet{Kob06} examined chromospheric velocity oscillations in sunspot 
penumbrae, and concluded that RPWs are the observational signature 
of slow-mode waves propagating along expanding magnetic field lines. 
Employing two-dimensional mapping of power and oscillatory period, 
\citet{Rou03} were also able to suggest that RPW phenomena could 
be attributed to near-acoustic field-aligned upwardly-propagating waves. 
\citet{Blo07} also provided momentum to the interpretation that RPWs are simply a 
chromospheric signature of upwardly-propagating acoustic waves by 
combining high-resolution spectra with Fourier phase-difference analysis 
to reveal how they readily propagate 
along magnetic fields in a low $\beta$ (i.e. dominated by 
magnetic pressure) environment. 
In the solar 
photosphere, the strong magnetic field strengths associated with sunspot 
umbrae result in a large magnetic pressure \citep[i.e. $\beta \ll 1$;][]{Mat04}. In this 
regime, the magnetic field is not influenced by the motion of the plasma, and as 
a result tends to a potential configuration of minimal energy \citep{Sol93, BorIch11}. 
However, the magnetic field strength decreases radially away from the sunspot 
umbra, rapidly causing the gas pressure to dominate towards the edge of 
the penumbra \citep[i.e. $\beta > 1$;][]{Gar01, Pus10}. 

The field of magneto-hydrodynamic (MHD) seismology has rapidly risen to 
the forefront of solar physics research in recent years. Such techniques 
allow the physical quantities and structures of sub-resolution solar 
features to be derived directly from the analysis of waveforms propagating 
through the local plasma 
\citep[][to name but a few]{Uch70, Rob84, Ver04, McE06, Ban07, Van07, Van08}. 
The advent of modern, high-resolution ground- and space-based 
instrumentation has allowed MHD seismology to uncover a wide range of 
new solar phenomena in the lower solar atmosphere, including torsional 
Alfv{\'{e}}n waves \citep{Jes09}, sausage waves \citep{Mor11}, 
oscillations in the magnetic field \citep{Fuj09, Moreels13}, the presence 
of dual oscillating modes \citep{Mor12}, and aspects of mode 
conversion \citep{Jes12b}. By studying the fundamental parameters of 
ubiquitous MHD wave motion in the lower solar atmosphere, we are 
provided with a key opportunity to probe the constituent mechanisms 
behind some of the most dramatic and widespread solar phenomena.

In this paper, we utilise high spatial and temporal resolution 
observations to investigate the propagation 
characteristics of RPWs in the lower solar atmosphere. 
We employ Fourier analysis and potential magnetic field 
extrapolations, in a previously unexplored way, 
to study the variation of RPWs
in the vicinity of a near-circular sunspot, and relate the derived 
characteristics to the 
spatially dependent magnetic field inclination angles 
of the underlying sunspot.

\begin{figure*}
\epsscale{1.0}
\plotone{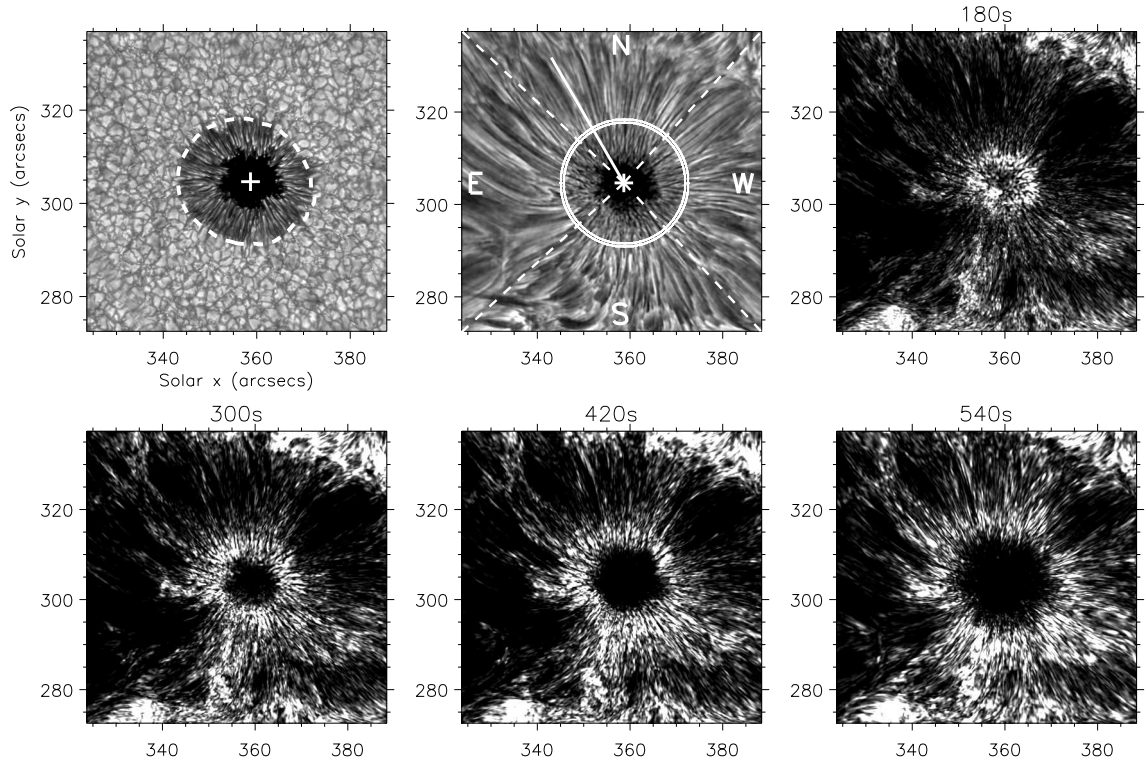}
\caption{Simultaneous images of the blue continuum (photosphere; upper left) and 
H$\alpha$ core (chromosphere; upper middle), acquired at 
$16$:$44$~UT on 2011 December 10. A white cross marks the 
barycenter of the sunspot umbra, while a white dashed line in the 
continuum image displays the extent of the photospheric $\beta = 1$ isocontour.
White concentric circles overlaid 
on the chromospheric image depict a sample annulus used to extract 
wave characteristics as a function of distance from the umbral 
barycenter, while the solid white line extending into the north quadrant 
reveals the slice position used for the time--distance 
analysis displayed in Figure~{\ref{time_distance}}. The dashed 
white lines isolate the active region into four distinct regions, 
corresponding to the N, W, S, and E quadrants. 
The scale is in heliocentric co-ordinates, where $1{\arcsec}\approx725$~km.
The remaining panels display a series of chromospheric power maps 
extracted through Fourier analysis of the H$\alpha$ time series, indicating 
the locations of high oscillatory power (white) with periodicities equal to 
$180$, $300$, $420$, and $540$~s. 
As the period of the wave becomes longer, it is clear that the 
location of peak power expands radially away from the umbral barycenter. 
\label{images}}
\end{figure*}

\section{Observations}
\label{obs}
The observational data presented here are part of a sequence obtained during 
$16$:$10$ -- $17$:$25$~UT on 2011 December 10, with the Dunn Solar 
Telescope (DST) at Sacramento 
Peak, New Mexico. The Rapid Oscillations in the Solar 
Atmosphere \citep[ROSA;][]{Jes10b} and newly-commissioned 
Hydrogen-Alpha Rapid Dynamics camera \citep[HARDcam;][]{Jess12} multi-wavelength 
imaging systems were employed to image a 
location surrounding active region NOAA $11366$, positioned at heliocentric co-ordinates 
($356${\arcsec}, $305${\arcsec}), or 
N$17.9$W$22.5$ in the conventional 
heliographic co-ordinate system. 
ROSA continuum 
observations were acquired through a 52{\,}{\AA} bandpass filter centered 
at 4170{\,}{\AA}, and employed a common plate scale of $0{\,}.{\!\!}{\arcsec}069$ 
per pixel, providing a diffraction-limited field-of-view size 
of $69{\arcsec}\times69{\arcsec}$. 
HARDcam observations employed a $0.25${\,}{\AA} filter centered on the 
H$\alpha$ line core (6562.8{\,}{\AA}), and utilised a spatial 
sampling of $0{\,}.{\!\!}{\arcsec}138$ per pixel, providing a 
field-of-view size ($71{\arcsec}\times71{\arcsec}$) 
comparable to the ROSA continuum image sequence.
During the observations, high-order adaptive optics \citep{Rim04} 
were used to correct wavefront deformations in real-time. The acquired images were 
further improved through speckle reconstruction algorithms \citep{Wog08}, 
utilizing $64 \rightarrow 1$ and $35 \rightarrow 1$ restorations for the 
continuum and H$\alpha$ images, resulting in reconstructed cadences of 
$2.11$~s and $1.78$~s, respectively. 
Atmospheric seeing conditions remained excellent 
throughout the time series. However, to insure accurate co-alignment between the 
bandpasses, broadband time series were Fourier co-registered and de-stretched using a 
$40 \times 40$ grid, equating to a $\approx$$1{\,}.{\!\!}{\arcsec}7$ 
separation between spatial samples 
\citep{Jes07}. Narrowband images, including those from HARDcam, were corrected 
by applying destretching vectors established from simultaneous broadband reference 
images \citep{Rea08, Jes10a}. Sample images, incorporating all image processing steps, 
can be viewed in Figure~\ref{images}.

The Helioseismic and Magnetic Imager \citep[HMI;][]{Sch12} onboard the 
Solar Dynamics Observatory \citep[SDO;][]{Pes12} 
was utilised to provide simultaneous vector magnetograms of active 
region NOAA~$11366$. The Milne-Eddington vector magnetograms 
were provided with a cadence of 
$720$~s and incorporate a two-pixel spatial 
resolution of $1{\,}.{\!\!}{\arcsec}0$. 
In addition, one 
contextual HMI continuum ($6173${\,}{\AA}) image, acquired at $16$:$10$~UT, was 
obtained for the purposes of co-aligning the HMI data with images of the 
lower solar atmosphere. 
The HMI data were processed using the standard {\tt{hmi\_prep}} 
routine, which includes the removal of energetic particle hits. 
Subsequently, $200${\arcsec} $\times$ $200${\arcsec} 
sub-fields were extracted from the processed data, 
with a central pointing close to that of the ground-based image 
sequences. Using the HMI continuum 
context image to define absolute solar co-ordinates, our ground-based 
observations were subjected to cross-correlation techniques to provide sub-pixel 
co-alignment accuracy between the imaging sequences. 
To do this, the plate scales of our ground-based observations 
were first degraded to match that of the HMI continuum 
image\footnote{Data analysis was performed on full-resolution 
({\rmfamily i.e.} non-degraded) image sequences}. Next, squared mean absolute 
deviations were calculated between the 
datasets, with the ground-based images subsequently 
shifted to best align with the HMI reference image. 
Following co-alignment, the maximum x- and y-displacements 
are both less than one tenth of an HMI pixel, or $0{\,}.{\!\!}{\arcsec}05$ 
($\approx$$36$~km).

\begin{figure*}
\epsscale{0.8}
\plotone{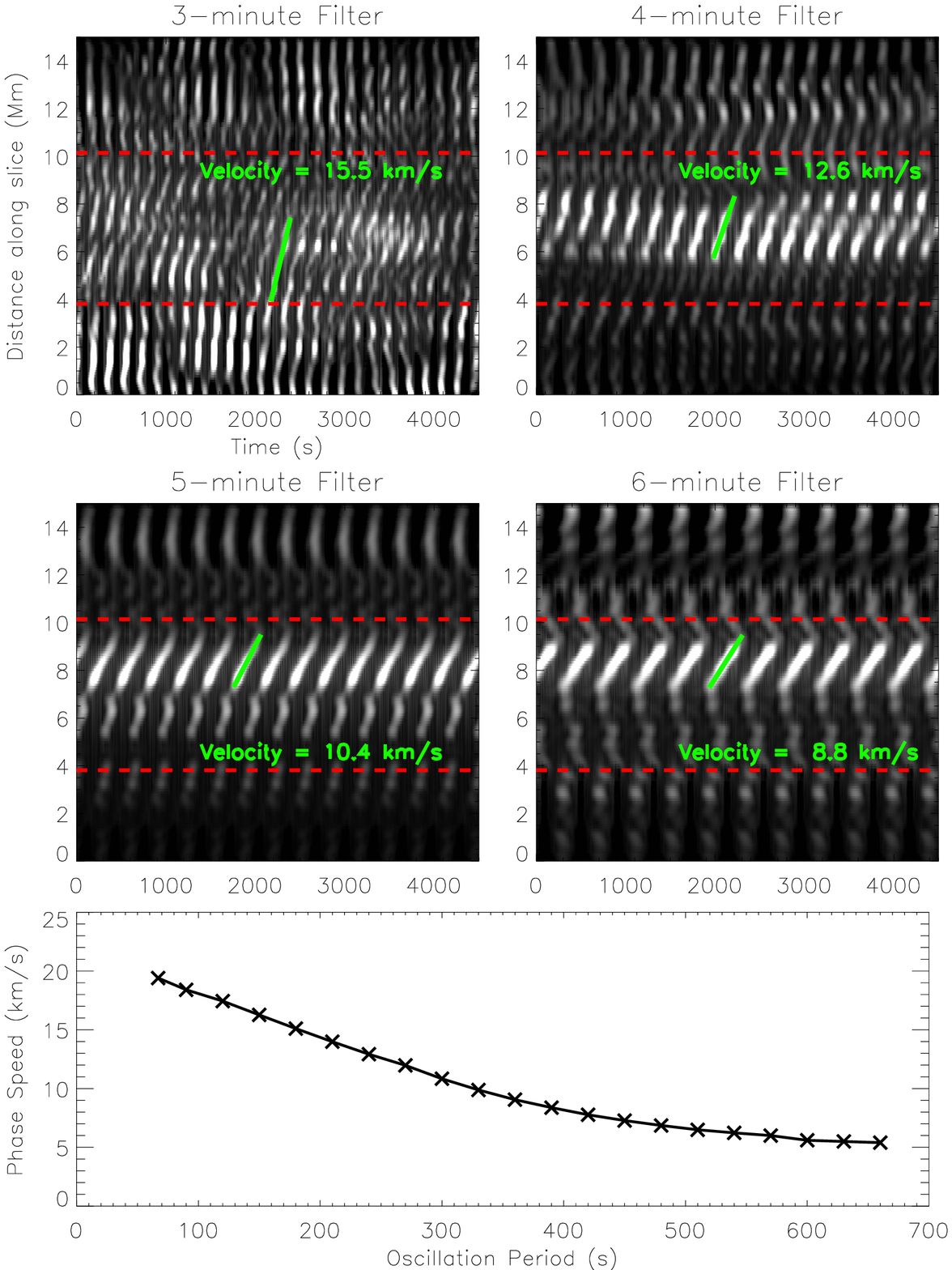}
\caption{Time--distance diagrams corresponding to the $3$ (upper-left), 
$4$ (upper-right), $5$ (middle-left), and $6$~(middle-right)~minute 
Fourier filtered time series, where the y-axis corresponds to the 
spatial extent occupied by the slice overlaid in the 
upper-middle panel of Figure~{\ref{images}}. Red 
horizontal dashed lines highlight the inner- and outer-penumbral 
boundaries at $\approx$$3.8$ and $\approx$$10.1$~Mm, respectively, 
from the umbral barycenter. Solid green lines highlight the lines-of-best-fit 
used to calculate the period-dependent phase speeds. Each time--distance 
diagram consists of $150$ spatial ($\approx$$15$~Mm) by 
$2528$ temporal ($75$~minutes) pixels$^{2}$. The lower panel displays 
the RPW phase speed (in km{\,}s$^{-1}$) as a function of 
oscillatory period. 
\label{time_distance}}
\end{figure*}

\section{Analysis and Discussion}
\label{analy}
During the two hour duration of the observing sequence, no large scale eruptive 
phenomena (GOES A-class or above) were observed from the active region 
under investigation. Examination of a time-lapse movie of HARDcam H$\alpha$ 
images revealed clear and distinctly periodic outflows along the chromospheric 
canopy in a direction away from the sunspot, consistent with RPW phenomena
\citep{Giov72, Zir72}. The irregular shapes of 
typical sunspot structures, often coupled with a non-continuous penumbral 
structure, result in convoluted wave patterns at increasing distances from the 
umbral edge \citep{Alis92, Tzio06}. However, the nearly circularly symmetric nature 
of our observed sunspot revealed a wealth of propagating wave fronts at 
distances far exceeding the outer penumbral edge. 

To verify the presence of wave phenomena in our H$\alpha$ 
field-of-view, we employed a Fourier-based 
filtering algorithm to isolate and re-generate new time series which 
had been decomposed into frequency bins, 
each seperated by $30$~s, and 
corresponding to periodicities of $1$--$11$~minutes. 
Each Fourier filter bin was peaked at $60$, $90$, 
$120$, \dots, $660$~s, and incorporated a Gaussian-profiled window 
spanning a half-width half-maximum of $\pm$$15$~s. 
We must stress that our 
Fourier filtering algorithm 
acts only as a frequency ($\omega$) filter. Our algorithm does not make 
any assumptions regarding the spatial scales associated with a particular 
frequency, and as a result performs no filtering on the spatial 
wavenumber ($k$). 
Examination 
of the filtered time series clearly revealed that oscillatory signatures 
with longer periodicities occurred at increasing distances from the 
sunspot umbra, and propagated with diminishing phase speeds. To 
quantify these characteristics, time--distance analysis was employed 
\citep[see, e.g.,][]{DeM02, Jess12}. 
A sample slice position is displayed as a solid white line extending in the 
north-east direction of the H$\alpha$ sunspot depicted in 
Figure~{\ref{images}}. The resulting time--distance diagrams, for the 
filtered time series corresponding to the $3$, $4$, $5$, and $6$~minute 
Fourier windows, are displayed in the upper and middle panels of 
Figure~{\ref{time_distance}}. The 
propagating oscillatory signatures, visible as recurring white and black 
diagonal bands, clearly show how longer-period oscillations occur 
more dominantly at further distances from the sunspot umbra (marked 
as $0$~Mm on the y-axis). Importantly, these propagating waves appear 
to be continuous in time, indicating they are driven from a regular and 
powerful source. Furthermore, the gradient of the diagonal 
bands provides an indication of the 
horizontal (i.e. parallel to the solar surface) 
phase speed of the propagating waves. A 
line-of-best-fit reveals phase speeds of $\approx$$15.5$, 
$\approx$$12.6$, $\approx$$10.4$, and 
$\approx$$8.8$~km{\,}s$^{-1}$ for the $3$, $4$, $5$, and $6$~minute 
Fourier windows, respectively. Employing the line-of-best-fit technique to each 
of the other Fourier-filtered time series reveals phase speeds spanning 
$\approx$$18.5$~km{\,}s$^{-1}$ ($60$~s periodicity) to 
$\approx$$5.6$~km{\,}s$^{-1}$ ($660$~s periodicity). 
Interestingly, the range of measured phase velocities was independent of 
the orientation of the time--distance slice. Regardless of the azimuthal 
direction in which the one-dimensional slice was placed, the observed 
wave periods propagated with 
the same phase velocities. All of the measured 
phase velocities are plotted as a function of oscillatory 
period in the lower panel of Figure~{\ref{time_distance}}. 
While the characteristics and trends displayed by the 
RPWs studied here are consistent with those measured in previous studies 
\citep[e.g.,][]{Bris97, Kob04}, the implementation of 
Fourier filtering algorithms significantly improves the accuracy of the 
phase velocity measurements. Each pixel in the unfiltered time series 
consists of the superposition of a large number of independent wave 
forms, each with different periodicities and phase speeds. The filtered 
time series removes this ambiguity, and allows the exact phase velocity to be 
reliably extracted as a function of period and distance from the sunspot 
umbra.

\begin{figure*}
\epsscale{0.8}
\plotone{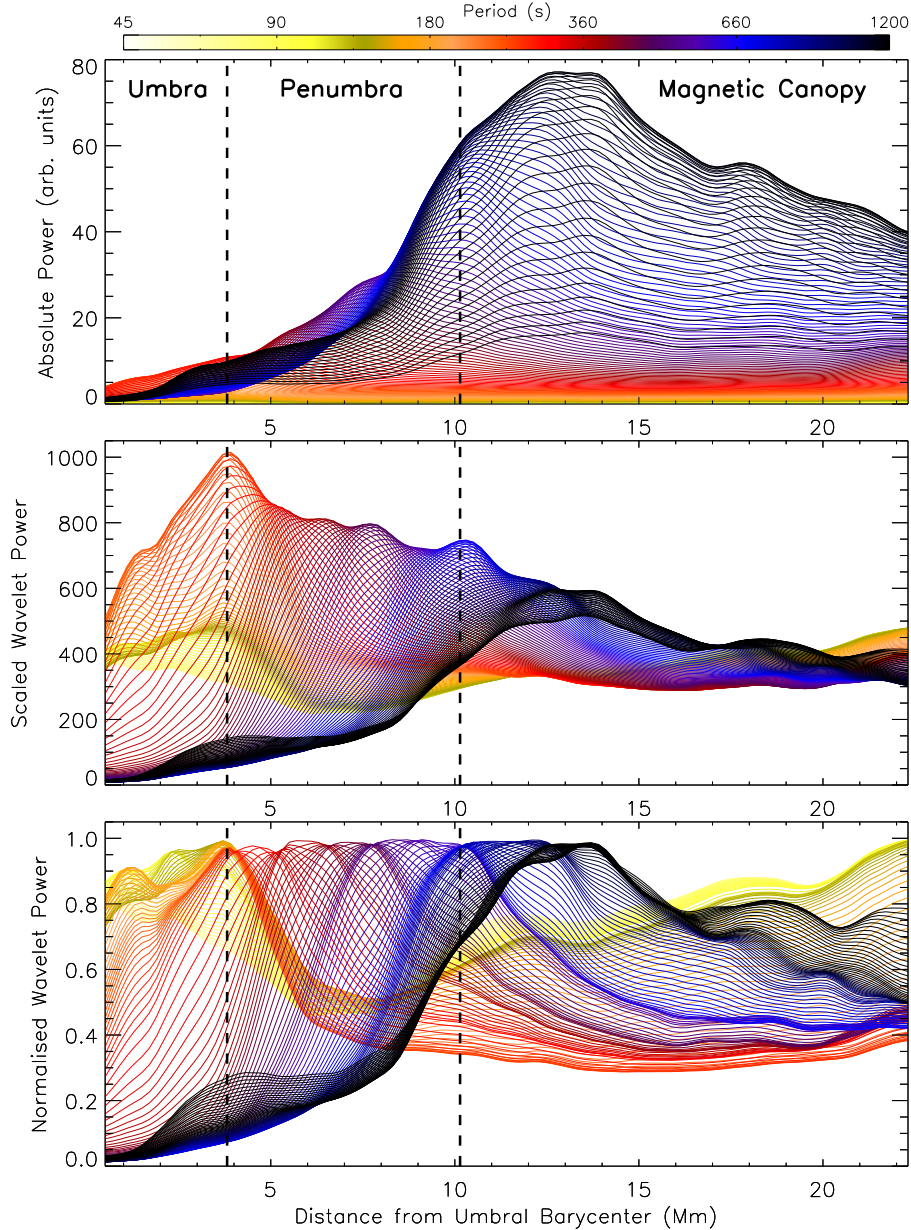}
\caption{{\it{Top:}} The azimuthally-averaged absolute Fourier power displayed as a 
function of radial distance from the umbral barycenter. 
{\it{Middle:}} The power spectra taken from the 
top panel and subjected to normalisation by the average power for that periodicity 
within the entire field-of-view. Thus, the vertical axis represents a factor of 
how much each period displays power above its spatially- and temporally-averaged 
background. {\it{Bottom:}} Power spectra normalised to their own respective maxima. 
Vertical dashed lines represent the radial extent of the umbral and penumbral 
boundaries, while the graduated color spectrum, displayed at the very top, 
assigns display colours to a series of increasing periodicities 
between $45$ -- $1200$~s.
\label{power_plots}}
\end{figure*}

To more closely quantify the characteristics associated with the propagating waves, 
Fourier-based analysis techniques were implemented on the unfiltered data. 
Due to the continual nature of the RPWs over the duration of our $75$~minute 
time series, we employed the methodology 
of \citet{Jes07} to apply strict Fourier analysis to 
the entire H$\alpha$ time series, allowing the signatures of wave phenomena to be 
easily extracted as a function of period. The upper-right and lower panels of 
Figure~{\ref{images}} display two-dimensional 
power maps for periodicities equal to $180$, $300$, $420$, and $540$~s. 
As from the time--distance analysis shown in Figure~{\ref{time_distance}}, it 
is clear that maximum Fourier power (displayed as bright white) for longer 
periodicities occurs at further radial 
distances from the sunspot umbra. Due to the nearly circularly symmetric nature 
of the sunspot under investigation, the best approach to study oscillatory 
behaviour as a function of radial distance is to examine the Fourier power over 
a series of expanding annuli. However, a location to use as the center of the 
expanding series of annuli must first be specified. To do this, continuum images of 
the sunspot umbra were isolated from the surrounding plasma. A time-averaged 
continuum image was created by averaging all 4170{\,}{\AA} blue continuum 
images over the entire 75~minute duration of the dataset. Next, 
the umbral pixels were defined as those with an intensity below 
$45$\% of the median granulation intensity. Features brighter than this were 
discarded, producing an accurately defined umbral perimeter, containing 
$\sim$$22{\,}000$ umbral pixels, or an area of $2.2 \times 10^{8}$~km$^{2}$. 
From this 
point, the umbral center-of-gravity, or `barycenter', was established, which formed 
the central co-ordinates of the annuli used in subsequent analysis. 
Here, a width of 5~pixels for each annulus was chosen, 
and subsequent annuli were spaced by 2~pixels from the preceding annulus. 
Some overlap between adjacent annuli was chosen to provide continuous 
radial coverage of the measured parameters, while still maintaining high pixel 
numbers to improve statistics. 
This was deemed essential, since at smaller radii from the umbral barycenter 
the number of pixels enclosed within a particular annulus is greatly reduced. 
Therefore, either a thicker annulus could be used, or more overlap could be provided 
between adjacent annuli. A thicker annulus was not desirable as this may artifically mask 
any changes in wave behaviour with distance. As a result, we chose to use an overlap between 
adjacent annuli, which greatly assists with measurements at small distances from the 
umbral barycenter. This overlap was kept constant for all annuli, regardless of the 
distance from the umbral barycenter. 
A sample annulus, which 
is 100~pixels ($\sim$$10$~Mm) from the umbral 
barycenter, is displayed in the upper-middle panel of Figure~{\ref{images}}.

\begin{figure*}
\epsscale{0.75}
\plotone{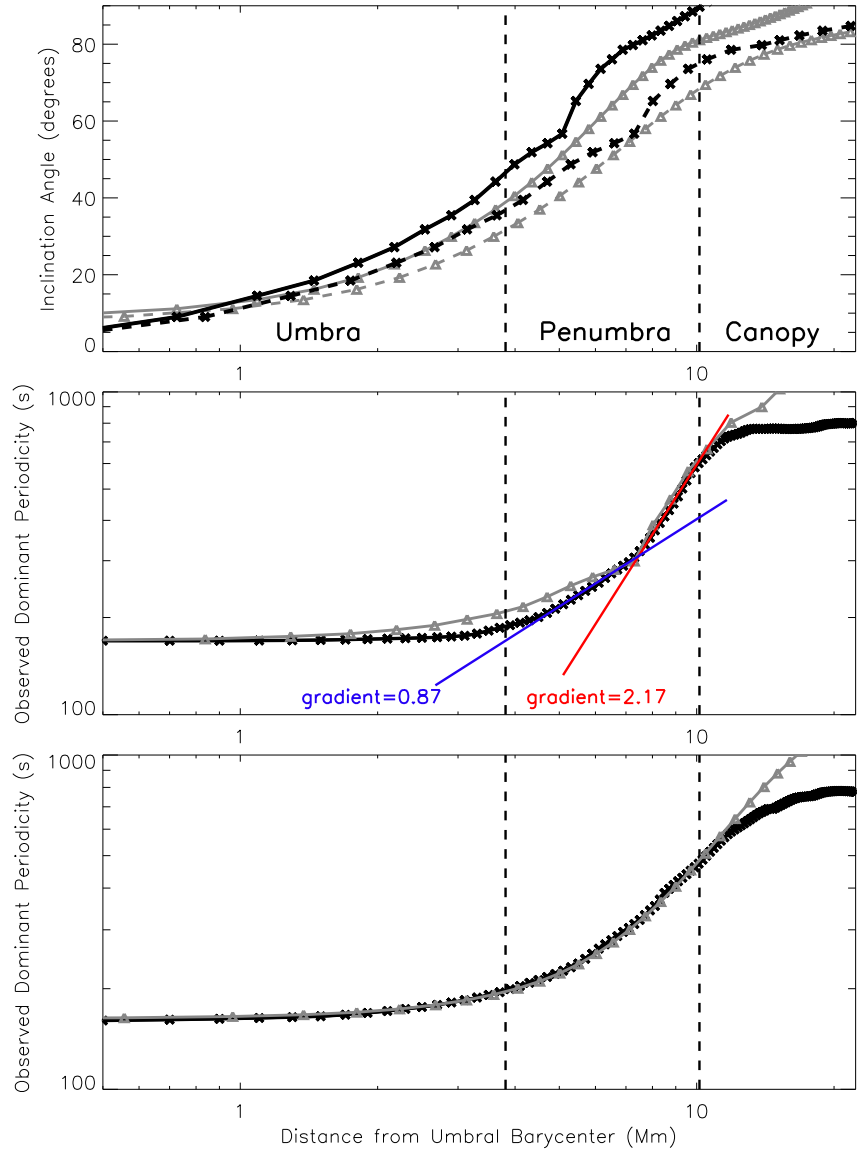}
\caption{The average inclination angles for the N (solid black line) and 
W+S+E (solid grey line) quadrants are plotted as a function of photospheric distance from the 
umbral barycenter in the upper panel. The dashed black and grey lines 
correspond to the same inclination angles, only plotted as a function of chromospheric 
distance from the umbral barycenter determined through magnetic field 
extrapolations. 
The observed dominant periodicity for the N sunspot quadrant is 
displayed in the middle panel on a log--log scale as a function of the radial distance away 
from the umbral barycenter (solid black line). 
Within the confines of the penumbra, there appear to 
be two distinct gradient slopes, as indicated by the blue and red lines 
of best fit. A solid grey line displays the acoustic cut-off period 
determined from the magnetic field inclination angles present in this quadrant 
of the sunspot. The lower panel displays the same information, only 
for the values averaged over the remaining three 
quadrants (W, S, and E). Vertical dashed lines indicate the inner- and 
outer-penumbral boundaries at $\approx$$3.8$ and 
$\approx$$10.1$~Mm, respectively. 
\label{dominant_period}}
\end{figure*}

Examination of the power contained within the annuli reiterated that higher-frequency 
intensity oscillations were confined to the inner penumbral edge, while lower-frequency 
perturbations were dominant at further increasing distances from the 
umbral barycenter (Figure~{\ref{power_plots}}). 
The highest-frequency intensity oscillations detected had a period 
$\approx$$45$~s, as defined by a $95$\% confidence level,
calculated by multiplying the power in the 
background spectrum by the values of 
$\chi^{2}$ corresponding to the 95$^{\mathrm{th}}$ 
percentile of the distribution \citep{Tor98, Mat03, Jes07}. 
In this case, the background spectrum is assumed to be consistent 
with pure photon noise (i.e. normally distributed in the limit of 
large number statistics), and following a $\chi^{2}$ distribution with 
two degrees of freedom. 
The longest-period intensity oscillations 
found, which superseded the $95$\% confidence criteria, were 
$\approx$$1200$~s, significantly under the Nyquist period of $2250$~s. The 
upper-panel of Figure~{\ref{power_plots}} reveals the absolute Fourier power, 
which has been 
spatially and temporarily averaged over each individual annulus, as a function 
of period and distance from the umbral barycenter. Here, long-period oscillations 
(displayed as darker colored lines) appear to display higher oscillatory power; a 
direct result of larger amplitudes accompanying the longer-period 
waves \citep{Did11}. 
Normalisation of the power spectra, by the average power 
contained within the entire field-of-view for each corresponding periodicity, reveals 
more meaningful structural information (middle panel of Figure~{\ref{power_plots}}). In this 
plot, the vertical axis gives a direct representation of how much each periodicity 
displays power above its spatially- and temporally-averaged background. Here, 
the shorter-period waves ($45 - 180$~s) display their peak power at the 
umbral/penumbral edge, with their relative power approximately three 
orders-of-magnitude higher than the background. As distance from the umbral 
barycenter is increased, the peak power is dominated by further increasing 
periodicities, with a dominant period $\approx$$640$~s at the outer 
penumbral edge.

It is visually clear in Figure~{\ref{power_plots}} that the dominant 
periodicity increases significantly 
between the inner and outer penumbral edges. This effect has been 
attributed to the inclination angles of the magnetic field lines 
increasing as a function of distance from the sunspot umbra \citep{Bel77}. As 
a result, the acoustic cut-off becomes heavily modified, thus allowing 
the free propagation of longer-period waves at increasing distances 
from the sunspot umbra. \citet{Blo07} displayed clear evidence 
of this phenomena by comparing the wave power spectra with the 
dispersion relations presented by \citet{Cen06}. However, the authors 
employed single-slit spectroscopic measurements with a spatial 
resolution $0{\,}.{\!\!}{\arcsec}8$, and were therefore 
unable to examine all locations within the sunspot umbra to a high 
degree of precision. To exploit our high spatial resolution observations, 
and to quantify the change in period at increasing distances from the 
sunspot umbra, we calculated the dominant periodicity as a function of 
distance from the umbral barycenter for 4 distinct regions of the sunspot, 
consisting of the north (N), west (W), south (S), and east (E) 
quadrants (see upper-middle panel 
of Figure~\ref{images}). For each quadrant, the dominant periodicity was 
defined as the periodicity which had the most relative power within 
each annulus segment. It was found that the W, S, and E quadrants 
displayed a gradual change in the dominant period as a function of 
distance from the umbral barycenter (black line in the lower 
panel of Figure~\ref{dominant_period}). However, the N quadrant displayed two 
distinct gradients between the inner and outer penumbral edges when 
displayed on a log--log scale. 
From the inner penumbral edge ($\approx$$3.8$~Mm from the umbral 
barycenter), to $\approx$$7.4$~Mm from the umbral barycenter, a 
line-of-best-fit reveals a periodicity, $P$, equal to, 
\begin{equation}
P = (53.13 \pm 1.03){\,}D^{{\,}(0.87 \pm 0.01)} \ {\mathrm{,}}
\end{equation}
where $0.87$ is the gradient of the line-of-best-fit, $53.13$ is the 
periodicity when the distance from the umbral barycenter, $D$, 
equals unity,
and the errors listed are the $1$$\sigma$ uncertainty estimates provided 
by the fitting function. 
The second distinct gradient manifests between 
$\approx$$7.4$~Mm and the outer penumbral edge $\approx$$10.2$~Mm 
from the umbral barycenter. Here, a line-of-best-fit allows the 
periodicity to be expressed as,
\begin{equation}
P = (3.99 \pm 1.08){\,}D^{{\,}(2.17 \pm 0.03)} \ {\mathrm{.}}
\end{equation}

While the general increase in period as a function of distance from the 
sunspot umbra has been well documented in previous studies, this is the 
first observational evidence of two distinct gradients present in a 
resulting period--distance plot (middle panel of Figure~{\ref{dominant_period}}). 
Thus, two important 
questions are why is the N quadrant of the sunspot different from the others? 
And what physical mechanisms are responsible for this 
pronounced effect? To investigate, we examined the vector magnetic 
field information collected by the HMI instrument. The 
Very Fast Inversion of the Stokes Vector \citep[VFISV;][]{Bor11} 
algorithm was utilised to decompose the magnetograms into components 
parallel ($B_{x}$ and $B_{y}$) and perpendicular ($B_{z}$) to the solar 
surface. 
Specifically, the initial vector field components (filename 
`hmi.ME\_720s\_fd10' 
on the Joint Science Operations Center, or JSOC, catalog) provide three 
values for each pixel related to the field strength, inclination angle, 
and non-disambiguated azimuthal angle. Azimuthal disambiguation 
of the transverse magnetic field vectors was undertaken using the 
algorithms of \citet{Rud11}. The disambiguated vector magnetograms 
were subsequently transformed into line-of-sight, east-west, and 
north-south components (the so-called ``basic'' transformation) 
following the procedures outlined in the HMI users 
guide\footnote{Users guide available at 
ftp://pail.stanford.edu/pub/HMIvector/documents/vector\_guide.pdf}. 
Finally, the transformation from the line-of-sight co-ordinate system 
into the correct heliographic projection (called either the ``advanced'' 
or ``vertical/horizontal'' decomposition) was performed in accordance 
with \citet{Gar90}. 
While the VFISV Milne-Eddington inversion algorithm returns a single 
magnetic field component, there is no doubt 
that the true magnetic configuration 
will be more complicated than that portrayed by even the 
relatively high resolution HMI observations. Particularly, in sunspot 
penumbrae, individual flux tubes have been found to show an 
interlocking-comb configuration \citep{Wei04}, whereby weaker 
and stronger field components have different inclination angles 
over very short distances. 
In addition, the magnetic structure cannot be entirely current-free, and 
is therefore also highly dynamic and constantly evolving \citep{Tho08}. 
However, for now we use the single field component 
provided by HMI as a first approximation of the dominant spatially 
resolved magnetic field vectors present in our observations. 
The derived values allowed the absolute magnetic field strength, $B$, 
to be computed for each pixel within our field-of-view. 
Importantly, the absolute 
magnetic field strength provided us with the ability to estimate the location of the 
$\beta = 1$ layer in the solar photosphere. 
The plasma $\beta$ is traditionally defined as the ratio between the gas 
pressure ($P_{g}$) and the magnetic pressure ($P_{B} = B^{2}/8{\pi}$), 
where $B$ is the absolute magnitude of the magnetic field strength. 
The plasma gas pressure, $P_{g}$, 
was estimated using the \citet{Mal86} average sunspot model `M', with 
parameters used which corresponded to the HMI magnetogram formation 
height of $\sim$$300${\,}km \citep{Bru90, Nor06, Fle11}. A temperature, 
$T = 3400${\,}K, and hydrogen number density, 
$n_{H} = 1 \times 10^{16}${\,}cm$^{-3}$, 
obtained for an atmospheric height $\sim$$300${\,}km from \citet{Mal86}, 
were used to estimate 
$P_{g}$. The resulting plasma $\beta$ was calculated according to 
$\beta = P_{g}/P_{B} = 8{\pi}n_{H}Tk_{B}/B^{2}$, where $k_{B}$ is the 
Boltzmann constant. 
An isocontour corresponding to the photospheric $\beta = 1$ 
layer is overplotted using a dashed white line on the blue continuum 
image displayed in the upper-left panel of Figure~{\ref{images}}. 
As a result, all plasma inside the isocontour is $\beta \leqslant 1$, allowing 
the whole sunspot structure to be considered as a low-$\beta$ region. 

\begin{figure}
\begin{center}
\includegraphics[scale=.385]{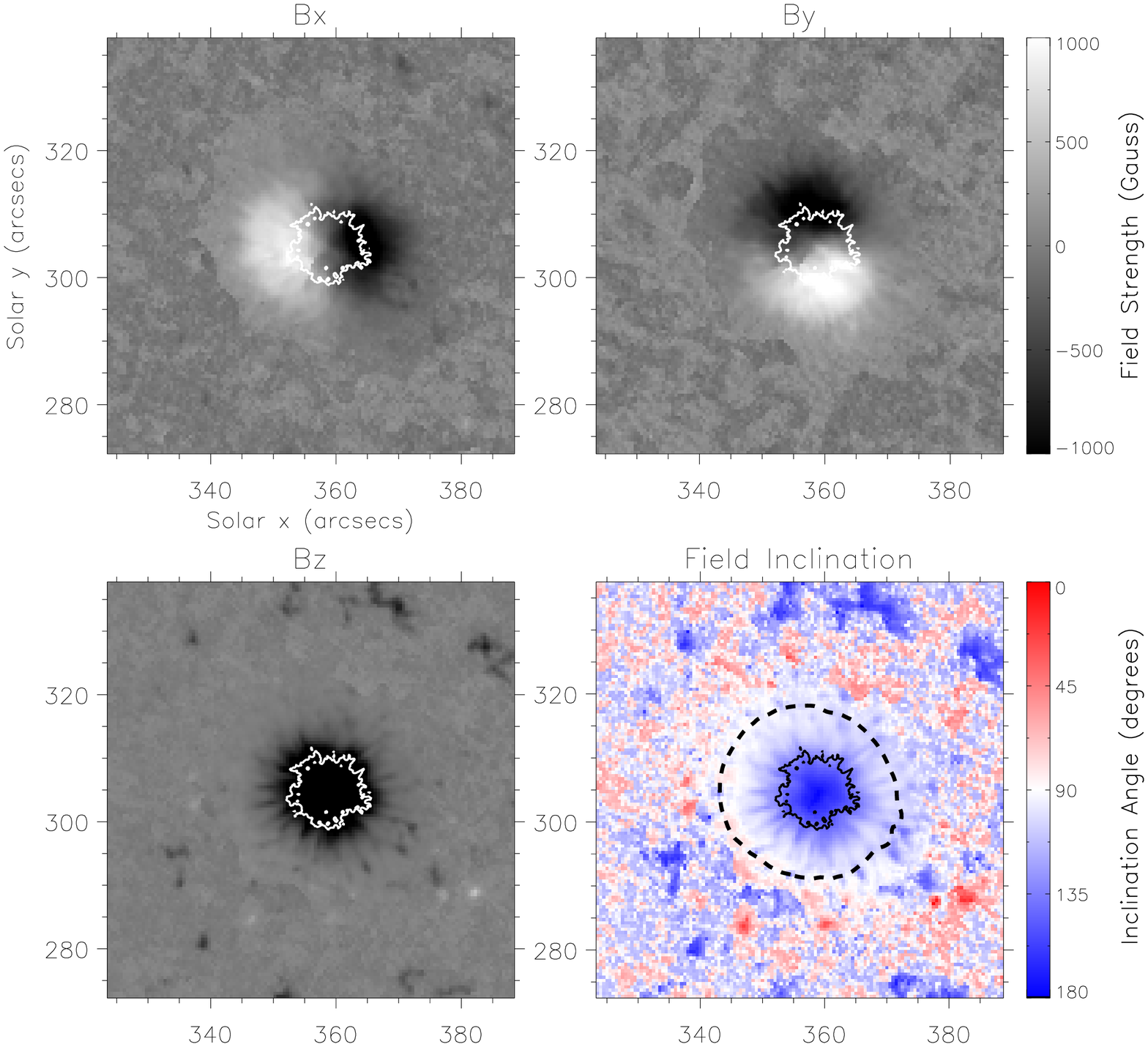} \\
\includegraphics[scale=.55]{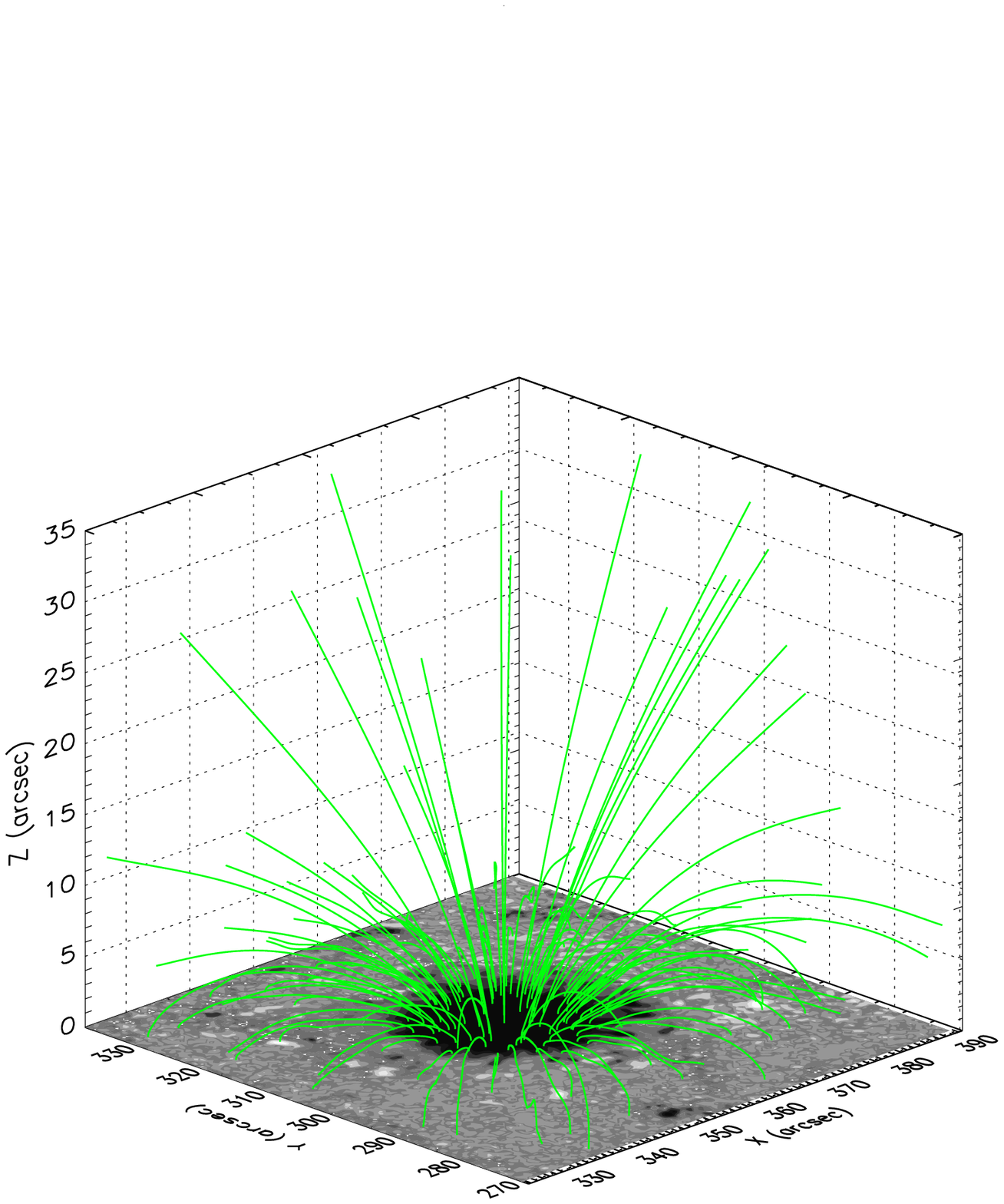}
\end{center}
\caption{HMI vector field components parallel 
($B_{x}$, upper left; $B_{y}$, upper right) and 
perpendicular ($B_{z}$, middle left) to the solar surface for 
NOAA~$11366$, where 
the magnetic field strength is displayed in greyscale, and 
artifically saturated at $\pm$$1000$~G to aid clarity. The 
middle-right panel reveals the inclination angles of the magnetic 
field vectors to the solar normal, while the lower panel displays 
the extrapolated magnetic fields (green lines) overlaid on the 
$B_{z}$ map, from an angle of $45$$\degr$ to 
the solar surface. The solid contours in each panel display the 
location of the inner penumbral boundary, while the dashed 
line in the middle-right panel displays the photospheric 
$\beta = 1$ isocontour 
which is representative of the outer penumbral boundary. 
The axes are displayed in heliocentric arcseconds.
\label{magnetogram}}
\end{figure}

To investigate the role that the magnetic field geometries may play 
in the creation of the period kink found in the N quadrant (solid black line 
in the middle panel of Figure~\ref{dominant_period}), we utilised the 
potential force-free field extrapolation code of \citet{Guo12}. 
Following the 
methods detailed in \citet{Met06} and \citet{Lek09}, magnetic field 
extrapolations were performed for the sunspot under investigation using 
the derived heliographic components as the initial conditions. 
Maps displaying the specific $B_{x}$, $B_{y}$, and $B_{z}$ magnetic components, 
in addition to the field vector angles for the observational field-of-view under 
investigation, are shown in the upper and middle panels of Figure~{\ref{magnetogram}}. 
Note that the inclination angles, $\theta$, displayed in the middle-right panel of 
Figure~{\ref{magnetogram}} are in the range $0-180${\degr}, where $180${\degr} 
corresponds to a negative polarity with a magnetic field vector pointing towards the 
center of the Sun \citep{Rez12}. For the purposes of displaying 
how the inclination angles vary as a function of radial distance, the 
true vector angles were recalculated so that $0${\degr} 
corresponds to a direction parallel to the solar normal (regardless of the absolute 
direction of the magnetic field vector), and $90${\degr} coincides with directions 
perpendicular to the solar surface. 
The azimuthally averaged inclination angles to the solar normal are displayed 
as a function of distance from the umbral barycenter for the N and W+S+E quadrants 
in the upper panel of Figure~{\ref{dominant_period}}. 
Importantly, the magnetic field extrapolations allow us to accurately determine 
what locations the HMI-derived inclination angles 
(middle-right panel in Figure~{\ref{magnetogram}}) correspond to at chromospheric 
heights. As the photospheric inclination angles increase as a function of distance 
from the umbral barycenter (solid black and grey lines in the upper panel of 
Figure~{\ref{dominant_period}}), one would expect that the chromospheric counterpart 
of these photospheric magnetic fields would lie at ever-increasing distances from the 
umbral barycenter. By tracing the photospheric magnetic field lines 
\citep[acquired at a formation height of $\sim$$300${\,}km;][]{Bru90, Nor06, Fle11} 
up to a chromospheric height corresponding to the core of the H$\alpha$ line 
\citep[$\sim$$1700${\,}km;][]{Ver81, Lee12}, we are able to display how the 
photospheric field inclinations would behave when displayed as a function of 
chromospheric distance. In order to maximise the accuracy of this procedure, the magnetic field lines were traced 
along incrementally-connected 3D vectors obtained via the potential magnetic field 
extrapolations, thus increasing the accuracy over a more-simplistic straight-line approximation 
between photospheric and chromospheric heights. 
Often the selection of a specific H$\alpha$ formation height is frought with 
difficulties, particularly in regions of high magnetic field strengths such 
as sunspots and pores. \citet{Lee12} have made considerable progress 
in recent years using full three-dimensional simulations of chromospheric 
plasma, including environments with a low plasma $\beta$ (i.e. 
dominated by magnetic pressure). These authors 
found an average H$\alpha$ formation height that varies between 
$1100 - 1900$~km depending on the local optical depth. 
Particularly, there is likely to be small chromospheric opacity in our 
observations of the umbral core, implying we are seeing further down into 
the photospheric layer \citep{Rut07}. From the H$\alpha$ image displayed 
in Figure~{\ref{images}}, it is clear that the sunspot umbra is dark and well 
defined, suggesting that the opacity is greatly reduced and that the 
formation height 
may be as low as $1100$~km. However, in the penumbra and 
chromospheric canopy regions, where we actually report on the 
characteristics of the RPWs, the opacity will be much higher in 
addition to the plasma $\beta$ being considerably larger. Thus, in 
these regions, the narrowband ($0.25$~{\AA}) H$\alpha$ core 
formation height is likely to be higher, and similar to the $\sim$$1700$~km 
proposed by \citet{Lee12}.

The upper 
panel of Figure~{\ref{dominant_period}} displays the HMI-derived inclination angles for 
both photospheric (derived directly from HMI observations; solid black and grey 
lines) and chromospheric distances (dashed black and grey lines). It is clear that 
when plotted on a chromospheric distance scale, the changes in the inclination 
angles occurs at further distances from the umbral barycenter when compared to 
the photospheric distance scale. Furthermore, a distinct jump in the inclination 
angle for the N quadrant is visible at a photospheric distance of $\approx$$5.1${\,}Mm, 
suggesting the inclination angle changes rapidly over a very short distance. 
The physical reason for this jump is unclear. Careful examination of the inclination 
map (middle-right panel of Figure~{\ref{magnetogram}}) 
reveals that a large fraction of the sunspot circumference displays prominent 
`spines' of low inclination angles radially extending 
outwards from the umbra. 
These spines are visible in the middle-right panel of Figure~{\ref{magnetogram}} 
as blue (i.e. more vertically orientated) structures extending radially outwards from 
the sunspot umbra. A more detailed examination of these spine structures 
reveals they preferentially correspond to bright penumbral filaments, as observed 
in both the ROSA and HMI continuum images. This link between vertical magnetic 
fields and bright penumbral filaments is in direct agreement with the work of 
\citet{Hof93} and \citet{Lan05,Lan07}. 
However, in the N quadrant these spines appear 
to be less pronounced and more curved, 
resulting in a more rapid, localised change in the magnetic field inclination angle 
as a function of distance from the umbral barycenter. 
When this inclination-angle kink is transferred to a chromospheric distance scale, it occurs at a distance 
of $\approx$$7.4${\,}Mm (dashed black line in the upper panel of Figure~{\ref{dominant_period}}); 
exactly the point where we observe a kink in the period--distance diagram for the RPWs. 
This technique allows us to directly compare the 
observed chromospheric wave phenomena with photospheric magnetic field inclination angles 
derived from HMI vector magnetograms.
The output of the magnetic field extrapolations can be viewed in 
the lower panel of Figure~{\ref{magnetogram}}. 
The near-potential configuration of 
the sunspot, in addition to its nearly circularly symmetric composition, results 
in magnetic field lines expanding outwards in all directions from the underlying 
umbra. 

Magnetic field inclination angles play an important role in the propagation of 
magneto-acoustic wave modes, since in the presence of gravitational 
fields they can only propagate upwards at frequencies above the 
acoustic cut-off, $f_{c}$, defined as,
\begin{equation}
f_{c} = \frac{g\gamma}{2{\pi}C_{s}} \ , \\
\label{cut_off_freq}
\end{equation}
where $C_{s} = \sqrt{{\gamma}RT / {\mu}}$ 
is the local sound speed, $T$ is the temperature, $\gamma$ is the 
ratio of specific heats, $\mu$ is the mean molecular weight, and $R$ 
is the gas constant. It was predicted by \citet{Bel77}, that in regions of 
low-$\beta$ plasma an effective gravity, $g$, on a particular magnetic 
field line can be expressed as,
\begin{equation}
g = g_{0} \times cos{\,}{\theta} \ , \\
\end{equation}
where the gravitational acceleration, $g_{0}$, is decreased by the 
cosine of the inclination angle, $\theta$, with respect to the solar 
normal. 
From Equation~{\ref{cut_off_freq}}, it is clear that the maximum 
cut-off frequency is determined by the minimum values of $\theta$ 
(to maximise the numerator) and $T$ (to minimise the denominator) 
along the path of wave propagation. As a result, both parameters 
($\theta$ and $T$) should be taken at the level which corresponds 
to the minimum temperature along a particular field line; the 
``cut-off height''. We assume that both the inclination angle and the 
temperature will increase beyond the cut-off height, and as a result, 
the cut-off height will define the frequency spectrum of waves that 
can propagate beyond the temperature-minimum level. 
According to the atmospheric sunspot models of \citet{Mal86}, the 
temperature-minimum region is located $280 - 530$~km above 
the continuum optical depth $\tau_{500{\mathrm{nm}}} = 1$. 
Here, we adopt the cut-off height as that corresponding to the 
formation height of the HMI magnetograms ($\sim$$300$~km).

The spatially-dependent cut-off frequencies were calculated 
using a minimum temperature, $T=3400$~K, 
${\mu} = 1.3$~g{\,}mol$^{-1}$, $g_{0} = 274$~m{\,}s$^{-2}$ 
(and then combined with the extrapolated $cos{\,}{\theta}$ 
terms to find the spatially-dependent effective gravity), 
$R = 8.31$~J{\,}mol$^{-1}${\,}K$^{-1}$, and 
${\gamma} = 5/3$ for an ideal monoatomic gas. 
The azimuthally averaged cut-off periods for the N and W+S+E 
sunspot quadrants are displayed as a function of chromospheric distance from the 
umbral barycenter using grey lines in 
the middle and lower panels of Figure~{\ref{dominant_period}}, respectively. 
The kink present in the inclination angle (N quadrant; $\approx$$5.1${\,}Mm 
photospheric distance or $\approx$$7.4${\,}Mm chromospheric distance)
diagram in the upper panel of Figure~{\ref{dominant_period}} results in a 
similar kink in the corresponding cut-off period (grey line in the middle panel 
of Figure~{\ref{dominant_period}}). Contrarily, the more smoothly varying 
inclination angles present in the W+S+E quadrants results in an equally smooth 
variation in the resulting cut-off period (grey line in the lower panel of 
Figure~{\ref{dominant_period}}). It is clear that the 
observed dominant periodicities 
closely follow the acoustic cut-off period, thus remaining 
consistent with the work of \citet{Bel77}, and highlighting the importance of 
being able to accurately relate chromospheric phenomena to the underlying 
photospheric magnetic field geometry. 

From the structure of a sunspot, it is clear that RPWs observed at increasing 
distances from the umbral barycenter will have travelled along magnetic field 
lines that are more inclined to the solar normal. This has direct consequences 
on the acoustic cut-off period, and as a result, longer 
periods are found at increasing distances from the umbral barycenter, 
producing a gradual change 
in the observed period when the magnetic field inclination angles are slowly varying 
(see, e.g., the black line in the lower panel of Figure~{\ref{dominant_period}}). 
However, 
when a more-rapidly varying magnetic field geometry is present, 
such as that corresponding to the N region of the 
sunspot under investigation, the resulting period--distance diagram (middle panel 
of Figure~{\ref{dominant_period}}) clearly 
reflects this magnetic complexity by revealing a distinct kink in the resulting 
relationship. 
Here, a rapid, localised inclination of the magnetic field lines results in a faster increase in the 
dominant periodicity due to the reduced cut-off frequency, $f_{c}$. 
The fact that the observed wave periods can be accurately constrained by the 
magnetic field inclination angles and the associated cut-off period, implies, at least 
for the sunspot under investigation, that its low-$\beta$ nature allows it to 
be accurately modelled using force-free magnetic field 
extrapolations, ultimately providing the ability to directly relate chromospheric phenomena 
to the underlying photospheric magnetic field complexity. 
Interestingly, the magnetograms displayed in Figure~{\ref{magnetogram}} 
also shine light on the intricate structuring present in the H$\alpha$ 
image displayed in the upper-middle panel of Figure~{\ref{images}}. 
It appears that the H$\alpha$ canopy structure has termination points 
visible in the north-west and south-east corners of the field-of-view. 
Figure~{\ref{magnetogram}} clearly reveals that these regions contain 
pockets of high unipolar magnetic field strength, which are often 
the locations of concentrated groups of magnetic bright points 
\citep{Jes10a}. In 
addition, these features typically give rise to 
localized regions of high oscillatory power (see the high oscillatory 
power present in the north-west and south-east corners of the 
Fourier power maps displayed in Figure~{\ref{images}}), which 
explains the presence of such oscillatory phenomena in these 
corners of the field-of-view. 
The termination of H$\alpha$ canopy structures and the 
manifestation of high oscillatory power are consistent with previous 
studies of magnetic bright point groups \citep{Jes09, Jes12b, Law11}.

We also suggest that the reverse approach can be implemented 
to obtain key information on the magnetic field geometries surrounding 
sunspots. 
In such a ``reverse'' regime, the presence of a kink in a 
period--distance diagram, the precise location of the kink, in addition to the 
gradients before and after the kink can all be used to provide valuable 
structural information related to the underlying magnetic field geometry. 
By establishing the permitted wave periods, the spatial 
variance of the magnetic field strengths, the cut-off periods,  
and thus the orientation of the underlying magnetic fields 
can be estimated directly. 
Importantly, the derived relationships 
between the underlying magnetic field geometries connecting the photosphere 
to the chromosphere, and the characteristics of RPWs observed in the upper 
chromosphere, directly supports the interpretation of \citet{Chri00, Chri01}
that these phenomena are the chromospheric signature of 
upwardly-propagating magneto-acoustic waves generated in the photosphere.

\section{Concluding Remarks}
\label{conc}
We have combined one of the highest resolution solar instruments currently 
available with a series 
of detailed computations to show that the observed variations in 
RPW periods can be explained by the influence of the low-frequency acoustic 
cut-off period. 
In the vicinity of the sunspot, the dominant periods 
are determined by the acoustic cut-off, which is directly influenced by 
the local inclination angles of the magnetic fields to the solar normal. Waves 
which pass this filtering selection criteria
are then allowed to be channelled into the chromosphere where they are 
observed as RPWs. 
A kink in the period--distance relationship for RPWs may be found if the magnetic 
field geometry changes rapidly over a relatively short 
distance. This kink can be more 
or less pronounced depending on the field inclination angles, and their 
associated rate of change, present in a 
particular locality. 
Importantly, this 
means that the spatial distribution of dominant wave periods directly reflects 
the magnetic geometry of the underlying sunspot, thus opening up a wealth of 
possibilities in future MHD-seismology studies. 

Using dedicated Fourier filtering algorithms, we 
have accurately measured the period-dependent phase velocity of RPWs. 
The 
shortest period waves ($\sim$$60${\,}s) have a phase velocity of 
$\approx$$18.5$~km{\,}s$^{-1}$, while the longest period waves ($\sim$$660${\,}s) 
demonstrate a phase velocity of 
$\approx$$5.6$~km{\,}s$^{-1}$. 
In addition, the intrinsic relationships we 
find between the underlying magnetic field geometries connecting the photosphere 
to the chromosphere, and the characteristics of RPWs observed in the upper 
chromosphere, directly supports 
the interpretation that these phenomena are the chromospheric signature of 
upwardly-propagating magneto-acoustic waves generated in the photosphere. 

\acknowledgments
D.B.J. wishes to thank the European Commission and the Fonds 
Wetenschappelijk Onderzoek (FWO) for the award of a Marie Curie 
Pegasus Fellowship during which this work 
was initiated, in addition to the UK Science and Technology 
Facilities Council (STFC) for the 
award of an Ernest Rutherford Fellowship which allowed the completion 
of this project. 
The research carried out by V.E.R. is partly supported by grant 
MC FP7-PEOPLE-2011-IRSES-295272.
T.V.D. acknowledges funding from the Odysseus Programme of the FWO Vlaanderen 
and from the EU's 7th Framework Programme as an ERG with grant number 
276808. P.H.K. and D.H.M. are grateful to STFC for research support.
This research has been funded by the Interuniversity Attraction Poles 
Programme initiated by the Belgian Science Policy Office (IAP P7/08 CHARM). 
All authors would like to thank the anonymous referee for their helpful comments 
and detailed knowledge which significantly improved this manuscript.

{\it Facilities:} 
\facility{Dunn (HARDcam, ROSA)}, 
\facility{SDO (HMI)}.

\end{document}